\documentclass{article}
\usepackage{arxiv}
\usepackage[utf8]{inputenc} 
\usepackage[T1]{fontenc}    
\usepackage{hyperref}       
\usepackage{url}            
\usepackage{booktabs}       
\usepackage{amsfonts}       
\usepackage{nicefrac}       
\usepackage{microtype}      
\usepackage{lipsum}		
\usepackage{graphicx}
\usepackage{natbib}
\usepackage{doi}
\usepackage{amsmath}
\newcommand{\scs}{\scriptscriptstyle}

\title{Orbital rotation of spheroidal Mie particles driven by counter-propagating circularly-polarized beams}

\author{ E.~N. Bulgakov\\
	Kirensky Institute of Physics\\
 Federal ResearchCenter KSC SB RAS\\
	 50 Akademgorodok, Krasnoyarsk, 660036, Russia\\
	\And
	A.~E. Ershov \\
	Institute of Computational Modelling SB~RAS\\50 Akademgorodok, Krasnoyarsk, 660036, Russia\\
	IRC~SQC, Siberian Federal University \\ 79 Svobodny pr., Krasnoyarsk, 660041, Russia\\
		\And
		V.~Kimberg \\
	Theoretical Chemistry and Biology \\ KTH Royal Institute of Technology \\ SE-100 44 Stockholm, Sweden\\
		\And
V.~S.~Gerasimov \\
	Institute of Computational Modelling SB~RAS\\50 Akademgorodok \\
Krasnoyarsk, 660036, Russia \\
IRC~SQC, Siberian Federal University \\ 79 Svobodny pr., Krasnoyarsk, 660041, Russia\\
	\texttt{gerasimov@icm.krasn.ru} \\
		\And
	D.N.~Maksimov \\
	Kirensky Institute of Physics\\  Federal ResearchCenter KSC SB RAS\\
	 50 Akademgorodok, Krasnoyarsk, 660036, Russia\\
	 IRC~SQC, Siberian Federal University \\ 79 Svobodny pr., Krasnoyarsk, 660041, Russia\\
}

\date{}

\renewcommand{\shorttitle}{Optically Driven Orbital Rotation}

\hypersetup{
pdftitle={A template for the arxiv style},
pdfsubject={q-bio.NC, q-bio.QM},
pdfauthor={E.~N.~Bulgakov, A.~E.~Ershov, V.~Kimberg, V.~S.~Gerasimov, D.N.~Maksimov},
pdfkeywords={Orbital rotation, Mie particles, Optomechanics },
}

\begin{document}
\maketitle

\begin{abstract}
We theoretically consider orbital rotation of a spheroidal submicron particle in the field of two counter-propagating circularly polarized Gaussian beams. We derived equations connecting the parameters of the circular orbits centered on the beams axis to the optical force and torque. The equations show that, besides orbital rotation, the spheroidal particle simultaneously rotates around its equatorial axis. We found that two distinct dynamic regimes are possible. The orbital motion can be accompanied by a rapid proper rotation with angular velocity an order of magnitude larger than the angular velocity of the orbital rotation. Alternatively, the orbital and proper rotations can be synchronized. The direction of orbital rotation can either coincide with or be opposite to the direction of rotation of the electric vector. The findings are confirmed by direct numerical simulations. The results can be of use in development of nano-scale gyroscopes as
well in shape-selective sorting of submicron particles.
\end{abstract}


\section{Introduction}
Optical tweezers are instruments that allow for manipulating sub-micron particles by applying focused beams of light \cite{Polimeno2018, Pesce2020}. The control over a dielectric particle is achieved through the action of optical forces, which attract the particles to the beam hot-spot, if the permittivity of the particle is larger than that of ambient medium. 
Typically, in optical tweezers the optical gradient force \cite{Sukhov2017, Toftul2024a} restricts lateral motion by driving the particle to a stable equilibrium on the beam axis. Interestingly, this stable equilibrium is not the only possible scenario of the particle dynamics. Due to the non-conservative nature of the optical forces acting on a spherical Mie particle \cite{Sukhov2017} the axial equilibrium does not correspond to a minimum of a potential energy. Thus, under certain conditions, the dynamic system may exhibit a Hopf bifurcation to limit cycle attractors which correspond to stable orbital rotation of the particle around the beam axis \cite{Svak2018, Simpson2021}, whereas the longitudinal motion is absent since the counter-propagating beam have equal intensities. Thus far, the phenomenon of lateral orbital motion has been thoroughly investigated only with spherical particles \cite{Svak2018, Simpson2021, Brzobohaty2023}. At the same time, with non-spherical or anisotropic particles the dynamics may become even more involved due to coupling between rotation and translational degrees of freedom \cite{Yan2007, Xu2008, Simpson2010, Simpson2014, Chang2014, Jakl2014, Li2015, Qu2015, Li2016, Mitri2017}. In such a situation, a comprehensive description of the dynamics requires taking into account the optical torques \cite{He2025}. Once the symmetry of the particle is reduced \cite{Shi2020, Zhang2024} it may exhibit rotation \cite{Li2016, Reimann2018, donato2018optical, Nalimov2020, Edlund2022, Zielinska2023, zielinska2024long}, tumbling \cite{Mihiretie2012}, or oscillatory motion \cite{Mihiretie2014, Loudet2014, Hoang2016, Petkov2017, Arita2020}. Application of circularly or elliptically polarized beams results into transfer of spin angular momentum \cite{Bliokh2015} into proper rotation of non-spherical particle \cite{Brzobohaty2015, Kuhn2017a} paving a way to potential applications for micromachinines \cite{Kuhn2017, Bruce2020},  and enantioselective sorting \cite{Shi2020, Ali2024}. 


\begin{figure}[t!]
 \centering
\includegraphics[width=0.7\textwidth]{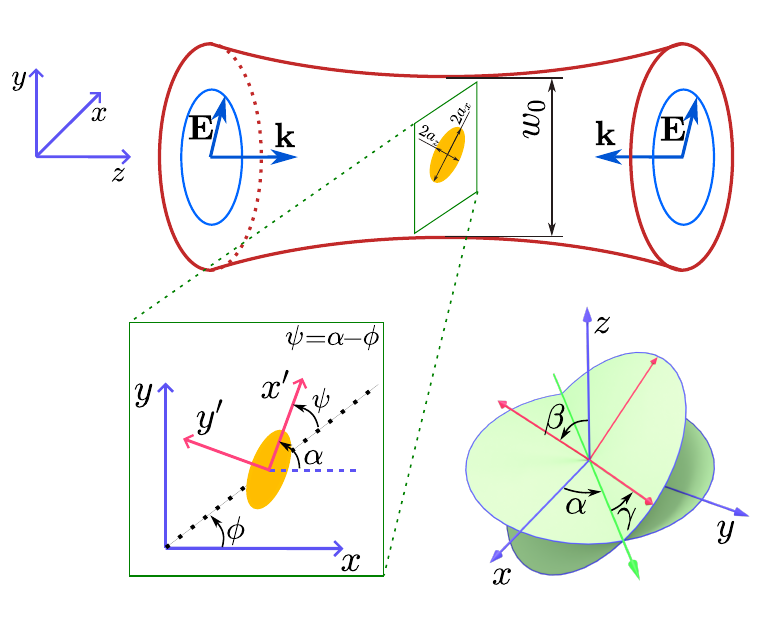}
\caption{ Spheroidal particle trapped by two counter-propagating coherent circularly polarized Gaussian beams. The sketch of the system is demonstrated on top. The coordinates in the $x0y$-plane are shown in the south-west corner, together with the definition of the rotating reference frame. In the south-east corner we illustrate the definition of the Euler angles.} \label{fig1}
\end{figure}

In this work, we theoretically investigate rotational motion of  spheroidal Mie particles trapped by coherent counter-propagating beams of opposite handedness. We assume that both beams are circularly polarized and synchronized in rotation of the electric vector. 
The geometric description of the set-up is presented in Fig.~\ref{fig1}.  In what follows we consider the spheroidal particle made of dielectric with $\epsilon=12$ and density $\rho=2~{\rm kg/m^3}$. The geometry of the particle is characterized by the aspect ratio $a_z/a_x=0.7$, where and $a_z$ is the length of the polar semi-axis of the spheroid, $a_x$ is the length of the equatorial semi-axis. The ambient medium is low vacuum with viscosity $\nu=4\cdot 10^{-7} ~\mathrm{Pa} \cdot {\rm sec}$. The paraxial Gaussian  beams are of wavelength $\lambda=1.55~{\mu}\mathrm{m}$ and have beam waist $w_0=5\lambda/\pi$. In our analytic approach we assume that the longitudinal motion is absent, i.e. the particle stays in the beam waist $z=0$. This corresponds to the maximal intensity hot spot along the $z$-axis. At the same time after setting the second Euler angle $\beta=\pi/2$ the polar axis of the spheroid is parallel to the $x0y$ plane and the system is symmetric to $z\rightarrow -z$. We assume that this symmetry is maintained during the motion and variables $z,~\beta,$ and $\gamma$ remain constant.  The configuration space, therefore, is spanned by three  dynamical variables $\alpha(t), ~x(t), ~y(t)$. 

\section{Forces and Torques} To recover the evolution of the dynamical variables  one has to account for both optical force and torque acting on the particles. This is done by calculating the force and the torque from the electromagnetic stress tensor with application of the standard formulas available in the literature \cite{Simpson2007, Neves2019}. The stress tensor, in its turn, is calculated from the full-wave numerically exact solution of the electromagnetic scattering problem obtained by the extended boundary condition method \cite{Barber1975, Kristensson2016}, which relies on field expansion into spherical vector harmonics. To integrate computation of the force and the torque with the extended boundary condition method, the optical field of the beams are also expanded into spherical vector harmonics  \cite{Doicu1997, Yan2007, Simpson2007}.
\begin{figure*}[b!]
\begin{center}
\includegraphics[width=\textwidth]{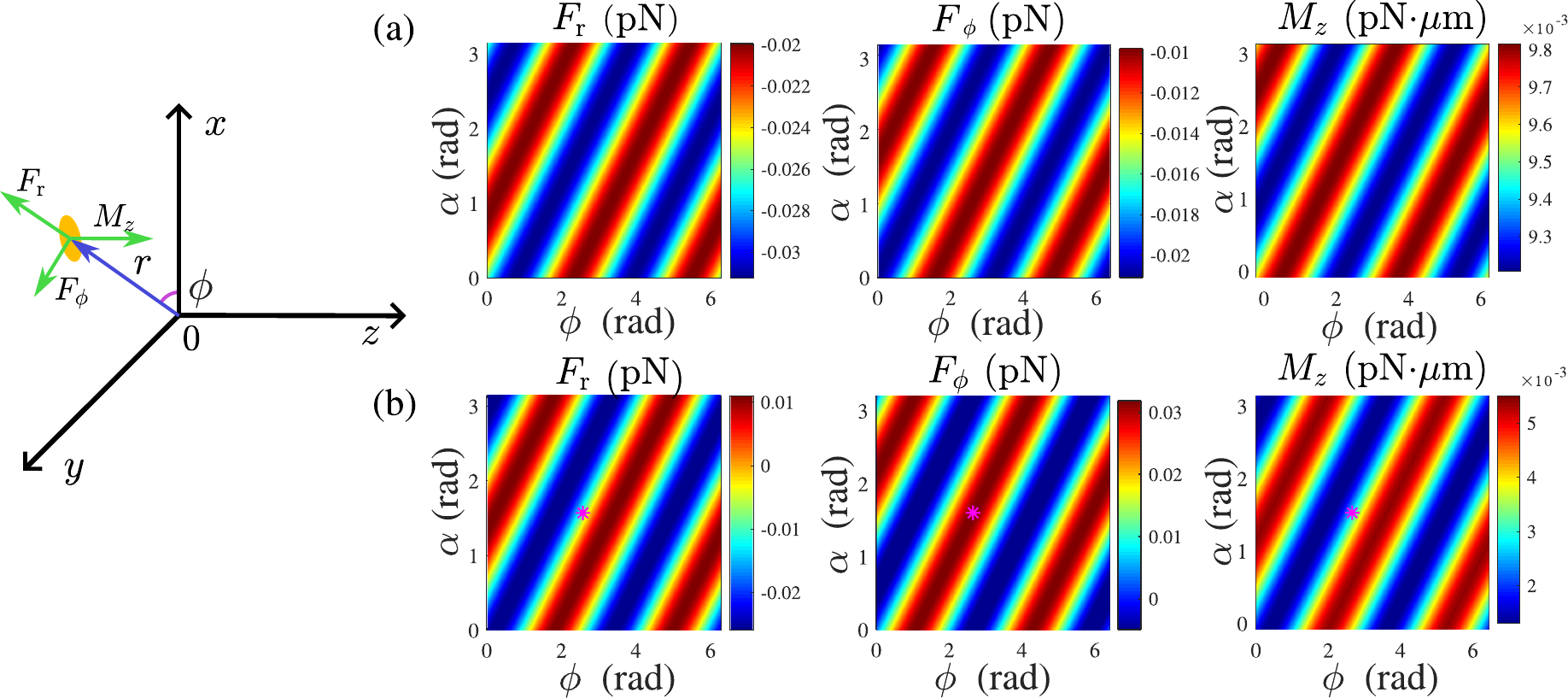}
\caption{ Optical force and torque on a spheroidal optical particle, $a_z/a_x=0.7$, subject to to the polarized  counter-propagating Gaussian beams shown in Fig.~\ref{fig1} as function of the first Euler angle $\alpha$ and azimuthal angle $\phi$; $z=0,~\beta=\pi/2$. The particle is located at the distance $r=1~{\mu\rm{m}}$ from the axis of the beams. The incident power of either beam is $1~\rm{mW}$. (a) $a_x=250~\mathrm{nm}$, and (b) $a_x=450~\mathrm{nm}$. The star on subplot (b) corresponds to $\alpha$ and $\phi$ in the solution Eq.~\eqref{data3}.
}\label{fig2}
\end{center}
\end{figure*}
It is worth mentioning that for a spherical particle there are two non-zero components of the net optical force in the cylindrical coordinates, namely $F_{\rm{r}}(r)$ and $F_{{\phi}}(r)$ whereas the $z$-component of the torque is zero, $M_z=0$. In the case of a spheroidal particle we expect a more involved picture with the first Euler angle $\alpha$ coming into play.
In Fig.~\ref{fig2} we illustrate the forces and torques acting on a spheroidal Mie particle in the set-up from Fig.~\ref{fig1} at the distance $r=1~\mu{\rm m}$ from the beams axis. One can see in Fig.~\ref{fig2} that all quantities, the force and the torque, depend on the difference
\begin{equation}\label{1}
\psi=\alpha-\phi,
\end{equation}
between $\alpha$ and $\phi$, see Fig.~\ref{fig1}, because there is no preferred direction in the $x0y$ plane in case of circular polarization.  
The dynamics under the action of the force and the torque from Fig.~\ref{fig2} will be studied both analytically and numerically. In our numerical simulations the optical and mechanical problems are solved simultaneously by taking into account the optical feedback onto mechanical degrees of freedoms. The numerical simulation of the mechanical problem is performed by simultaneously solving Newton's second laws of motion and Euler's equations by LSODE solver~\cite{Hindmarsh1983} taking into account viscosity according to \cite{Perrin1934}.

\section{Dynamics} We start with writing down the dynamic equations for a spheroidal particle driven by circularly polarized beams.
The components of the Stokes force in the rotating reference frame, see Fig.~\ref{fig1}, can be written as
\begin{equation}\label{2}
\begin{split}
    F^{\scs{\rm{(s)}}}_{x'}=-\nu  C_1 v_{x'}, \\
    F^{\scs{\rm{(s)}}}_{y'}=-\nu  C_3 v_{y'},
\end{split}
\end{equation}
where $v_{x'}$, $v_{y'}$ are the Cartesian components of velocity in the rotating reference frame, and $C_1$, $C_3$ are the shape-dependent drag correction factors \cite{Perrin1934}. 
After applying Eq.~\eqref{1} and Eq.~\eqref{2}, Newton's second law of motion in the 
cylindrical coordinates takes the following form
\begin{align}\label{Newton}
     & m(r\ddot{\phi}+2\dot{r}\dot{\phi})=F_{\phi}(r,\psi)-
    r\dot{\phi}{\nu}\left[C_1\sin^2(\psi)+C_3\cos^2(\psi)\right]-\dot{r}\frac{\nu}{2}(C_1-C_3)\sin(2\psi),
    \nonumber \\
     & m(\ddot{r}-r\dot{\phi}^2)=F_{\rm{r}}(r,\psi)-
    r\dot{\phi}\frac{\nu}{2}(C_1-C_3)\sin(2\psi)-
    \dot{r}{\nu}\left[C_3\sin^2(\psi)+C_1\cos^2(\psi)\right],
\end{align}
where $m$ is the mass of the particle. The above equations have to be complemented with the equation describing rotation of the spheroid about the equatorial axis
\begin{equation}\label{4}
    I_{\scs{\perp}}\ddot{\alpha}=M_z(\psi,r)-\nu\Gamma_{\scs{\perp}}\dot{\alpha},
\end{equation}
where $I_{\scs{\perp}}$ is the corresponding component of the moment of inertia tensor, and $\Gamma_{\scs{\perp}}$ is a size- and shape-dependent coefficient   that determines the resistance of the medium to rotation \cite{Perrin1934}.

The condition $\dot{r}=0$ corresponds to rotation around the $z$-axis with orbital angular velocity
$\Omega=\dot{\phi}$ in a circular orbit of radius $r$.
Equation (\ref{Newton}) is, then, reduced to
\begin{align}\label{Newton2}
     & 0=F_{\phi}(r,\psi)-
    r\dot{\phi}{\nu}\left[C_1\sin^2(\psi)+C_3\cos^2(\psi)\right],
    \nonumber \\
     & mr{\Omega}^2=-F_{\rm{r}}(r,\psi)+
    r\dot{\phi}\frac{\nu}{2}(C_1-C_3)\sin(2\psi).
\end{align}
Assuming 
that the proper angular velocity is much greater that orbital angular velocity $\omega\gg\Omega$,
where $\omega=\dot{\alpha}$, one can find the approximate analytic solution of Eq.~\eqref{Newton2} by averaging over a rapidly changing variable $\psi=\omega t -\phi$. After the average Eq.~\eqref{Newton2} becomes
\begin{align}\label{Newton3}
     & 0=\bar{F}_{\phi}(r)-
    r\Omega\frac{\nu}{2}\left(C_1+C_3\right),
    \nonumber \\
     & mr{\Omega}^2=-\bar{F}_{\rm{r}}(r),
\end{align}
where $\bar{F}_{\phi}(r), \bar{F}_{\rm{r}}(r)$ are the  components of the net force averaged over $\psi$. There are two unknowns in Eq.~\eqref{Newton3}, namely,
the orbit radius $r$, and the orbital angular velocity $\Omega$. Equation (\ref{Newton3}) can be rewritten as a single equation below
\begin{equation}\label{Newton4}
\frac{\bar{F}^2_{\phi}(r)}{\bar{F}_{\rm{r}}(r)}=-\frac{r}{m}\left(\frac{\nu}{2}\right)^2(C_1+C_3)^2,
\end{equation}
meanwhile the angular velocities are given by
\begin{align}\label{velocities}
    & \Omega=\sqrt{\frac{|\bar{F}_{\phi}(r)|}{mr}}, \
    \omega=\frac{\bar{M}_z}{\nu\Gamma_{\perp}},
\end{align}
where $\bar{M}_z$ is the average $z$-component of the torque, and the expression for $\omega$ is derived from Eq.~\eqref{4}. Once the orbit radius is known the angular velocities can be found from Eq.~\eqref{velocities}.
Note that the left hand part of Eq.~\eqref{Newton4} is proportional to the power of the beams $P$ whereas the right hand part is independent of $P$. Thus, a decrease of the beam power results in shrinking of the orbit radius until $r=0$. By numerically solving Eq.~\eqref{Newton4} for a particle with $a_x=0.25~\mu{\rm m}$ we find that the collapse of the orbital motion occurs at power $P=2.45~\rm{mW}$. This can be interpreted a transition to the well-known scenario of a dielectric particle trapped on the beam axis.

Another possible scenario is the phenomenon of phase synchronization 
\begin{equation}\label{cynch}
    \omega=\Omega={\rm Const}.
\end{equation}
The phase-synchronization or phase-locking  is typically explained by coupling between two oscillatory degrees of freedom with the difference between individual natural frequencies smaller than the sum of coupling strengths \cite{strogatz2001nonlinear}. Then, the proper and orbital rotation can occur at the same compromise frequency, Eq.~\eqref{cynch}. 
The phase-synchronized solution can be obtained by applying Eq.~\eqref{cynch} to Eq.~\eqref{Newton2}. By setting
\begin{align}\label{10}
    & \Omega=\frac{M_z(\psi)}{\nu\Gamma_{\perp}}, \
     \psi={\rm Const}
\end{align}
and using Eq.~\eqref{10} in Eq.~\eqref{Newton2} one arrives at
\begin{align}\label{Newton5}
   & mr\left(\frac{M_z(\psi)}{\nu\Gamma_{\perp}}\right)^2\!=\!-F_{\rm{r}}(r,\psi)+
    r\frac{M_z(\psi)}{2\Gamma_{\perp}}(C_1-C_3)\sin(2\psi),     \nonumber \\
    & 0\!=\!F_{\phi}(r,\psi)-
    r\frac{M_z(\psi)}{\Gamma_{\perp}}\left[C_1\sin^2(\psi)+C_3\cos^2(\psi)\right].
\end{align}
Equation~\eqref{Newton5} can be solved numerically 
for parameters $\psi$ and $r$ that define phase-synchronized orbits.

\begin{figure*}[t!]
\centering
\includegraphics[width=\textwidth]{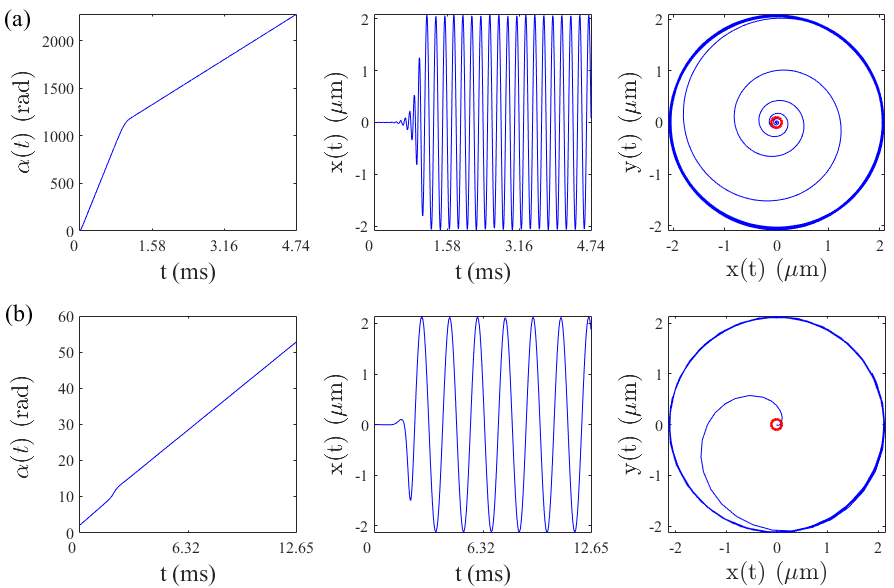}
\includegraphics[width=\textwidth]{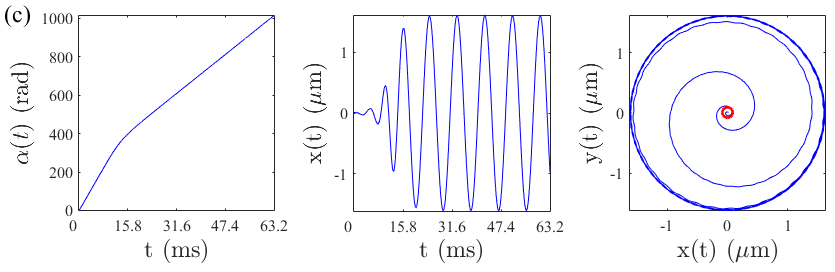}
\caption{ Rotation in a circular orbit. The first column shows the first Euler angle $\alpha$ against time, the second column -- the $x$-coordinate against time, the third column -- the trajectory in $x{0}y$ plane. The red circle in the right row  shows the initial condition. The initial kinetic energy is zero.  (a) The evolution of variables $\alpha, x, y$ in course of time for a smaller particle $a_x=0.250~\mu{\rm m}$ with a rapid proper rotation. The beam power $P=10~{\rm mW}$. (b) The phase synchronized trajectory for a larger particle  $a_x=0.450~\mu{\rm m}$. The beam power $P=0.45~{\rm mW}$. (c) Rapid rotation for $a_x=0.400~\mu{\rm m}$ with incident beam power $P=0.67~{\rm mW}$.}
\label{fig3}
\end{figure*}

\begin{figure*}[ht!]
\centering
\includegraphics[width=\textwidth]{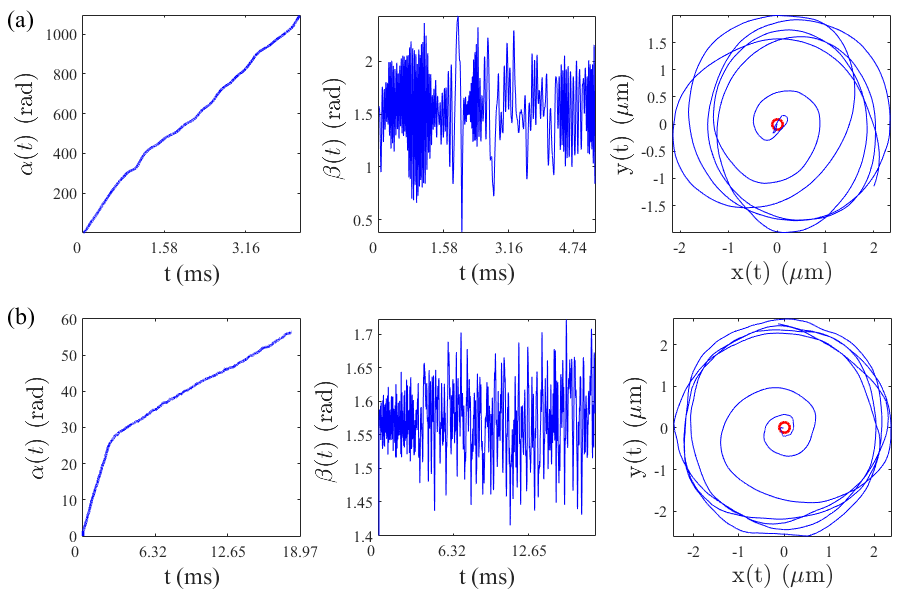}
\caption{ Effect of Brownian force. (a) Rapid proper rotation trajectory from Fig.~\ref{fig3}~(a) at $T=300~{\rm K}$. (b) Phase-synchronized trajectory from Fig.~\ref{fig3}~(b) at $T=6~{\rm K}$.
The mid-column shows the time dependence of the second Euler angle $\beta$.}
\label{fig4}
\end{figure*}
Our finding are verified in comparison against direct numerical simulations. In Fig.~\ref{fig3}~(a) we demonstrate the typical dynamic picture for a particle with $a_x=0.250~\mu{\rm m}$. One can see in Fig.~\ref{fig3}~(a) that in course of time the particle is attracted to a circular orbit limit cycle with a constant proper angular velocity. The theoretical predictions based on solving Eq.~\eqref{Newton4} produced the following parameters of the circular orbit limit cycle
\begin{equation}\label{data1}
    r\!=\!2.059~\mu{\rm m},~\omega\!=\!3.26\cdot 10^{5}~{\rm \frac{rad}{sec}},~\Omega\!=\!3.29\cdot 10^{4}~{\rm \frac{rad}{sec}}. 
\end{equation}
These numbers are to be compared with those extracted from the data shown in Fig.~\ref{fig3}~(a)
\begin{equation}\label{data2}
    r\!=\!2.060~\mu{\rm m},~\bar{\omega}\!=\!3.00\cdot 10^{5}~{\rm \frac{rad}{sec}},~\bar{\Omega}\!=\!3.19\cdot 10^{4}~{\rm \frac{rad}{sec}},
    \end{equation}
where $\bar{\omega}$ and $\bar{\Omega}$ are the numerically obtained angular velocities averaged over the period of the orbit.
One can see from Eq.~\eqref{data1} and Eq.~\eqref{data2} that our theoretical approach predicts the parameters of the limit cycle to a good accuracy. In Fig.~\ref{fig3}~(b) we show a phase synchronized trajectory a particle with $a_x=0.450~\mu{\rm m}$. As before we see that in the course of time the system evolves to a circular orbit limit cycle. The proper angular velocity is now {two orders} of magnitude smaller.   
Equation~(\ref{Newton5}) yields the following theoretical predictions for the parameters of the limit cycle
\begin{equation}\label{data3}
    r=2.15~\mu{\rm m},~\omega=\Omega=1.61\cdot 10^3~{\rm \frac{rad}{sec}}. 
\end{equation}
The numbers in Eq.~\eqref{data3} are in full compliance with the numerical data in Fig.~\ref{fig3}~(b).
Our numerical simulations show that the regime of rapid rotation for smaller-sized particles can be observed with size in the range $0.20-0.25~\mu m$. However, the decrease of the particle size from the case shown in Fig.~\ref{fig3}~(a) is accompanied with the growth of the threshold beam power, up to $P=1~{\rm W}$ at $a_x=0.20~\mu{\rm m}$. For larger particles the synchronization regime persists as the size decreased to $a_x=0.42~\mu m$ and also requires an increase of power up to $P=1.4~{\rm W}$. Below $a_x=0.420~\mu m$ the phase synchronized regime becomes unstable. However, the regime of rapid rotation can be also observed at much lower power $P=0.67~{\rm mW}$ for the particle size $a_x=0.4~\mu m$, see Fig.~\ref{fig3}~(c). 

Some comments are due on the direction of rotation. The direction of the proper rotation is always dictated by the handedness of polarization in accord to the spin angular momentum transfer. Interestingly, the directions of orbital rotation can be both clockwise and anticlockwise, as one can see from Fig.~\ref{fig3}. The orbital rotation occurs if $F_{{\rm r}}(r,\psi)<0$. At the same time, the orbital rotation direction is dictated by the sign $F_{{\phi}}(r,\psi)$ with the positive signs leading to anticlockwise rotation. As one can see Fig.~\ref{fig2}, both directions of orbital rotation are possible depending on the size of the particles. The value of $\alpha$ and $\phi$ found by solving Eq.~\eqref{Newton5} for trajectory in Fig.~\ref{fig3}~(b) is marked by star in Fig.~\ref{fig2}~(b) indicating anticlockwise rotation.

Let us now discuss the experimental feasibility of the dynamic regimes shown in Fig.~\ref{fig3}. In a realistic experimental set-up the regular dynamics can be distorted by Brownian motion. To access this effect we compare the mechanical energies of proper $E_{{\rm prop}}$ and orbital $E_{{\rm orb}}$ rotations against the thermal energy $k_{{\rm B}}T$, where $k_{{\rm B}}$ is the Boltzmann constant, at temperature $T=300~K$. For the rapid proper rotation shown in Fig.~\ref{fig3}~(a) we find
$E_{{\rm prop}}/k_{{\rm B}}T=25$ and $E_{{\rm orb}}/k_{{\rm B}}T=52$ so the regular dynamics dominate over Browninan motion even at room temperature. In the case of synchronised motion shown in Fig.~\ref{fig3}~(b) we found $E_{{\rm prop}}/k_{{\rm B}}T=0.01$ and $E_{{\rm orb}}/k_{{\rm B}}T=0.6$. Therefore, synchronization can only be observed at temperatures of the order of Kelvin. To confirm our estimates we ran numerical simulation of particle dynamics with account of Brownian force according to \cite{Ermak1978, Loewen1994, morillo2019brownian}. The results are presented in Fig.~\ref{fig4}. In Fig.~\ref{fig4}~(a) we demonstrate the regime of rapid proper rotation at $T=300~{\rm K}$, while Fig.~\ref{fig4}~(b) shows the regime of synchronized rotation at $T=6~{\rm K}$. One can see from Fig.~\ref{fig4} that our estimates comply with direct numerical simulations. Note that, the dynamics under the action of Brownian forces and torques is modelled in 3D space with account of all components of the force and the torque. In particular, this permits fluctuation of the second Euler angle, so the polar axis of the spheroid is not exactly parallel to the $x0y$ plane but rather fluctuates about $\beta=\pi/2$ as seen from the mid-column in Fig.~\ref{fig4}.

\section{Conclusion} We have investigated orbital rotation of a spheroidal particle in the field of two counter-propagating circularly polarized Gaussian beams. Similar to the earlier findings for spherical particles \cite{Svak2018, Simpson2021} we observe that with increase in intensity the particle departs from the beam axis evolving to a circular orbit limit cycle.
Thus, this work extends the results of Svak {\it et al} \cite{Svak2018}, who demonstrated that a spherical dielectric particle exposed to a circularly polarized beam in optical tweezers can depart from axial equilibrium and undergo circular motion around the beam axis, driven by a Hopf bifurcation that produces coherent thermally excited orbits. We advanced this line of research by theoretically analyzing the orbital rotation of spheroidal particles, explicitly accounting for the transfer of spin angular momentum into particle rotation \cite{Bliokh2015}. 
As the central result, equations connecting the parameters of the circular orbits to the optical force and torque are derived analytically. The equations show that, besides circular orbital rotation, the spheroidal particle also rotates around one of its equatorial axes which is parallel to the beam axis. We found that two distinct dynamic regimes are possible. For example, with particles $a_x=0.250~\mu m$, and $a_x=0.400~\mu m$ in size the orbital motion is accompanied by a rapid proper rotation with angular velocity an order of magnitude larger than the angular velocity of the orbital rotation. For particles with $a_x=0.450~\mu m$ the orbital and proper rotations of the particle are synchronised, i.e. occur with the same angular velocity. The direction of orbital rotation can either coincide with or be opposite to the direction of rotation of the electric vector. The regime of rapid proper rotation can be observed at room temperature. However, the synchronized rotation can only be observed at temperatures of several Kelvins. We believe that this low temperature effect is experimentally feasible in a view of of the results on millikelvin cooling of optically trapped microparticles \cite{Li2011}. We speculate that lowering the temperature threshold for synchronized motion constitutes an interesting subject for future research. In our analytic model we neglected longitudinal motion. This leaves us with an unanswered question of whether the reported two-dimensional dynamic regimes are stable with respect to perturbations in $z$ and $\beta$. To address this question the stability of the orbits shown in Fig.~\ref{fig3} is confirmed by performing full-fledged three-dimensional modelling in six dimensional configuration space with account of the Brownian force. Interestingly, we found that with a further increase of intensity the orbital motion bifurcates to three-dimensional attractors to be considered in the future studies. We believe that the result presented can be of use in development of nano-scale gyroscopes as well in shape-selective sorting of submicron particles that can rotate in different orbits depending on their size.

\bibliographystyle{plain}
\bibliography{refs} 
\end{document}